\documentclass[aps,prl,showpacs,superscriptaddress,twocolumn]{revtex4}
\usepackage{amsmath, amssymb}
\usepackage{graphicx}
\usepackage{dcolumn}
\usepackage{bm}
\newcommand{\wick}[1]{:\!{#1}\!:}

\begin{document}


\title{Dark Energy from Quantum Matter}



\author{Claudio Dappiaggi}
\author{Thomas-Paul Hack}
\email{thomas-paul.hack@desy.de}
\affiliation{II. Institut f\"ur Theoretische Physik, Universit\"at Hamburg,
D-22761 Hamburg, Germany.}
\author{Jan M\"oller}
\affiliation{Deutsches Elektronen-Synchrotron DESY, Theory Group, D-22603 Hamburg, Germany}
\author{Nicola Pinamonti}
\affiliation{Dipartimento di Matematica, Universit\`a di Roma ``Tor Vergata'', I-00133 Roma, Italy.}


\date{\today}

\begin{abstract}
We study the backreaction of free quantum fields on a flat Robertson-Walker spacetime. Apart from renormalization freedom, the vacuum energy receives contributions from both the trace anomaly and the thermal nature of the quantum state. The former represents a dynamical realisation of dark energy, while the latter mimics an effective dark matter component.  The semiclassical dynamics yield two classes of asymptotically stable solutions. The first reproduces the $\Lambda$CDM model in a suitable regime. The second lacks a classical counterpart, but is in excellent agreement with recent observations.
\end{abstract}

\keywords{keywords to be decided}
\pacs{04.62.+v, 95.36.+x, 95.35.+d}

\maketitle

\noindent{\em Introduction.} During the last decade, cosmological observations have established an ongoing phase of accelerated expansion in the recent history of our Universe \cite{ries98, sper03, eisn05, Amanu10, wmap10}. According to the $\Lambda$CDM concordance model of cosmology, the acceleration results from a modification of Einstein's equations by a cosmological constant ($\Lambda$) which has just begun to dominate the total energy density of the Universe. More generally, such a contribution is called dark energy. In addition, the Universe is widely believed to be filled with cold dark matter (CDM), a non-luminous and weakly interacting component responsible for the formation and growth of large scale structures \cite{zw37, ruf70, eina09}, which outweighs the contribution of baryonic matter by a factor of ten. Despite all efforts, the dark energy lacks a sound theoretical understanding so far. The purpose of this letter is to shed a new light on these issues from the fundamental point of view of quantum field theory on curved spacetimes.

Compared to earlier attempts on this task - see for example \cite{Star80, Ander83, Parker01, Shap09, Koksma09} - the results presented here benefit from recent progress on a quantization scheme which is intrinsically tied to non-trivial spacetimes \cite{BFV, HW01}, and represents a significant leap forward in our understanding of quantum field theory on curved backgrounds. Our aim is to apply these new techniques in a cosmological framework in order to address the issue whether dark energy and, if possible, dark matter originate from fundamental, non-interacting quantum fields.

\vskip .2cm

\noindent{\em Cosmological quantum fields.} We consider a four-dimensional flat Robertson-Walker (RW) spacetime,
whose metric
$
ds^2 = -dt^2+a^2(t) {d\vec{x}}^{\,2}
$
is completely determined by the scale factor $a(t)$. The dynamical evolution of the Universe in terms of $a(t)$ is to be obtained from a solution of the semiclassical Einstein equations,
\begin{equation*}
	G_{\mu\nu}=8\pi G \langle \wick{T_{\mu\nu}} \rangle_{\Omega}\,,
\end{equation*}
where $G_{\mu\nu}$ is the Einstein tensor and $G$ the gravitational constant. On the right hand side, one has to take the expectation value of the regularised matter stress-energy tensor in a Gaussian state $\Omega$. The latter is chosen to fulfil the so-called Hadamard condition which fixes the UV properties of $\Omega$ and assures the existence of a well-defined notion of normal ordering of fields, $:\cdot:$ \cite{HW01}.  The consistency requirement that the expectation value of $\langle\wick{T_{\mu\nu}} \rangle_{\Omega}$ should be covariantly conserved unavoidably leads to a term in the trace of $\langle\wick{T_{\mu\nu}} \rangle_{\Omega}$ which has no classical counterpart, the trace anomaly \cite{Duff77, Wald78, Moretti}. Given a collection of $N_0$ conformally coupled real scalar fields $\phi_i$, $i=1,...,N_0$, with mass $m_i$, $N_{1/2}$ Dirac spinors $\psi_j$, $j=1,...,N_{1/2}$ with mass $m_j$, and $N_1$ massless vector fields on a RW spacetime, the trace reads \cite{Chris78, Chris79, DFP, DHP}
\begin{eqnarray}
\langle\wick{T^\mu_{\phantom{\mu}\mu}}\rangle_{\Omega} = \alpha \left( R^{\mu\nu}R_{\mu\nu}-\frac{R^2}{3}\right)  +c_1(m) R   + c_2(m) +  \notag  \\ + c_3 \square R - \sum_{i=1}^{N_0}m^2_i \langle \wick{\phi^2_i} \rangle_{\Omega} -  \sum\limits_{j=1}^{N_{1/2}}m_j \langle \wick{\overline\psi_j\psi_j} \rangle_{\Omega}\,.\label{anomaly}
\end{eqnarray}
Here, $R_{\mu\nu}$ is the Ricci tensor and $R$ the scalar curvature, while $\alpha=(2880\pi^2)^{-1}(N_0+11 N_{1/2}+62 N_1)$. Moreover, $c_1(m)$ and $c_2(m)$ are linear combinations of powers of the masses. The corresponding mass coefficients, as well as the coefficient $c_3$ itself, are subject to a finite renormalization freedom, and hence free parameters of the theory. Note that any state-dependent quantity appears multiplied by a suitable power of the mass, so that massless fields contribute only to the purely geometric terms.

Observing the homogeneity and isotropy of our metric ansatz, the semiclassical Einstein equations can be rewritten in terms of two coupled inhomogeneous ordinary differential equations,
\begin{eqnarray}\label{ODE}
\frac{\dot{\rho}}{H}+4\rho=-\langle\wick{T^\mu_{\phantom{\mu}\mu}}\rangle_{\Omega}\,,\qquad
 H^2=\frac{8\pi G}{3}\rho\,,
\end{eqnarray}
where dots denote derivatives with respect to $t$. Here, $H\doteq\frac{\dot{a}}{a}$ is the Hubble function and we set $\rho\doteq \langle\wick{T_{00}}\rangle_\Omega .$ The quantum energy density $\rho$ is obtained from \eqref{ODE} up to a solution of the corresponding homogeneous equation, which is of the form $\rho= C_0 a^{-4}$. As we are ultimately interested in describing the late time evolution of the Universe, which is characterised by a low spacetime curvature, we shall employ an ``adiabatic'' approximation and discard terms in $\rho$ with time derivatives of $H$. Hence, the contribution of the purely geometric, state-independent part in \eqref{anomaly} to the quantum energy density reads $$\rho_\text{geom}=3\alpha H^4-3c_1(m)H^2-\frac{c_2(m)}{4}\;.$$
As $H\ll m$ for generic masses in the late cosmic history, we supplement the adiabatic approximation by discarding ${\cal O}(H^5)$ terms in $\rho$. We retain ${\cal O}(H^4)$ terms to capture the lowest order effects of curved spacetime quantum fields.

\vskip .2cm

\noindent{\em State dependence.} We consider every quantum field to be in a state $\Omega$ which fulfils an approximate KMS condition at some instant of time $t = t_0$ in the past [23]. Hence, $\Omega$ depends on a fixed temperature parameter T equal for every field. Moreover, we demand that $T\ll m$ and compute all terms to the lowest non-trivial order in $T/m$. We start with the conformally coupled Klein-Gordon field. Following \cite{DHP2, Thomas}, the two-point function of $\Omega$ reads
$$\langle\phi(x)\phi(y)\rangle_\Omega=
	\int_{\mathbb{R}^3}
	\left(
	\frac{\overline{\phi_{\vec k}(x)}\phi_{\vec k}(y)}{1-e^{-\beta k_0}}  + \frac{{\phi_{\vec k}(x)}\overline{\phi_{\vec k}(y)}}{e^{\beta k_0}-1}
	\right)\frac{d{\vec k}}{8\pi^3}\;,
	$$
where $k=|{\vec k}|$ and $\phi_{\vec k}(x)\doteq T_k(t)e^{i{\vec k}{\vec x}}$ are suitably normalized Klein-Gordon modes. Furthermore, $k_{0}\doteq(k^2+m^2a_0^2)^{1/2}$ and $a_0\doteq a(t_0)$, whereas $\beta$ is the inverse temperature. On account of our choice of $\wick{\cdot}$ and $\Omega$, and up to the renormalization freedom already present in \eqref{anomaly}, we find
$$
\langle\wick{\phi^2}\rangle_\Omega = \int_0^\infty\!\! \frac{k^2}{2\pi^2} \left( \frac{e^{\beta k_0}+1}{e^{\beta k_0}-1} |T_k(t)|^2 -\frac{a^{-2}}{\sqrt{k^2+m^2a^2}} \right) dk.
$$
Considering our adiabatic approximation, we compute $|T_k(t)|^2$ by expanding it in terms of $H$ and its time derivatives so that the resulting $\rho$ depends only on $H^n$, $n<5$. This yields
$$
|T_k(t)|^2= g_1 + g_2 H^2+g_3 \dot{H}
+g_4H^4 + g_5\dot H H^2 \,.
$$
Here, $g_i$ for $i=1,...,5$ are suitable functions of $a$, $k$, and $m$. Since  $|T_k(t)|^2$ fulfils an ordinary differential equation which descends from the Klein-Gordon equation, these functions can be determined recursively by taking the UV properties of $\Omega$ as an initial datum. As shown in detail in \cite{Thomas}, we thus obtain
$$
- m^2\langle\wick{\phi^2}\rangle_\Omega = \frac{H^4+7\dot{H}H^2}{240\pi^2}
-
\frac{l m T^3}{\pi^2\;a^3}\,,
$$
up to the renormalization freedom accounted for in \eqref{anomaly} and up to higher orders in $T/m$. Here, $l$ is a monotonically decreasing function of $a_0\beta m$ with $l(0)=\zeta(3)$.

For Dirac fields, the object of interest is the expectation value $\langle\wick{\overline\psi \psi}\rangle_\Omega$. As  explained in \cite{DHP2, Thomas}, and barring the renormalization freedom already present in \eqref{anomaly}, this quantity can be computed as
\begin{gather*}
\langle\wick{\overline\psi \psi}\rangle_\Omega =
\int_0^\infty \frac{k^2}{\pi^2} {(2k^2|Q_k|^2-1)}\frac{e^{\beta k_0}-1}{e^{\beta k_0}+1}\,dk\,  + \\
 \frac{m}{\pi^2 a^2 } \int_0^\infty\frac{k^2  dk}{\sqrt {k^2 + a^2 \left(m^2+\frac{R}{12} \right)} } -
\frac{R^2}{1152m\pi^2} + {\cal O}(R^3)\,,
\end{gather*}
where $Q_k$ are normalised Dirac modes. Recalling our adiabatic approximation, we expand $|Q_k|^2$ in terms of $H$ and its derivatives up to sufficient powers. This yields
$$
|Q_k|^2= G_1 + G_2 \dot{H} + G_3 H^2 + G_4 \dot{H}H^2 + G_5 H^4 \;,
$$
where again the $G_i$ depend on $a$, $m$ and $k$. By a procedure similar to the scalar case we obtain
\begin{eqnarray*}
-m\langle\wick{\overline\psi \psi}\rangle_\Omega =
\frac{-19 H^4-15\dot{H}m^2+17 \dot{H} H^2}{240\pi^2}-\frac{3LmT^3}{\pi^2\;a^3}\,,
\end{eqnarray*}
\noindent see \cite{Thomas} for the details. Here, $L$ is a function sharing the properties of $l$.

Collecting the state dependent terms of $\langle\wick{T^\mu_{\phantom{\mu}\mu}}\rangle_{\Omega}$ and computing the corresponding energy density, one finds
$$\rho_\text{state}=\frac{3}{2880\pi^2}\left(-N_0+19 N_{1/2}\right)H^4+c_4(m)\frac{T^3}{a^3}\,.$$
\noindent Here, $c_4(m)>0$ is a fixed and constant linear combination of the field masses, and derivatives of $H$ have been discarded in accord with our adiabatic approximation.

\vskip .2cm

\begin{figure}
\includegraphics[width=1\columnwidth]{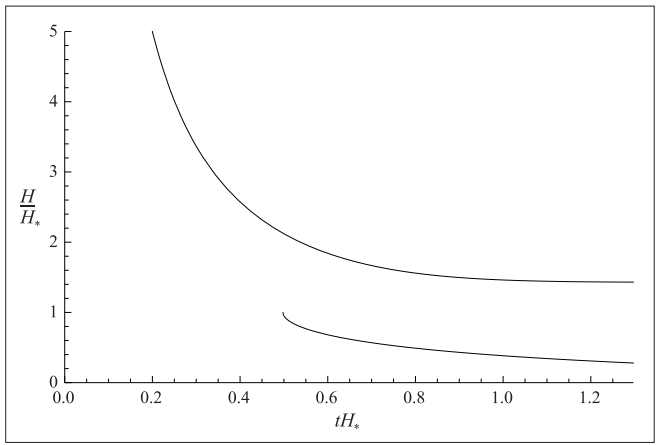}
\caption{Generic plot of the two branches. While $H_+$ displays a Big Bang singularity, $H_-$ becomes imaginary at finite time \cite{DHP}. The latter phenomenon is an artefact of the adiabatic approximation employed.}\label{Branches}
\end{figure}

\noindent{\em Effective Friedmann equation.} We insert $\rho = \rho_\text{geom}+\rho_\text{state}$ into the second equation in \eqref{ODE} to obtain an algebraic equation for $H$. If $N_{1/2}>0$, this equation is of fourth order in $H$. Its solutions may be interpreted as effective Friedmann equations and read
\begin{gather}
	H^2(a)_\pm=H_*^2
 \pm\sqrt{H_*^4 - \frac{C_1}{a^4} - \frac{C_2}{a^3} - C_3} \;,\label{effective}
\end{gather}
where $H_*$ depends on the number of fields and the renormalization freedom, $C_2$ depends on the field masses and $T$, $C_3$ is a free renormalization parameter, and $C_1$ is a multiple of the already mentioned integration constant $C_0$ present in $\rho$.  A qualitative analysis of \eqref{effective} displays two asymptotically stable fixed points, corresponding to de Sitter spacetime. Physical solutions approach one of these points with $H^2>0$ and $\dot{H}<0$, see figure \ref{Branches}. The value of the effective cosmological constant associated to the attractors is not fixed since it depends on the renormalization constants of the underlying quantum field theory \cite{DFP}.

\begin{figure}
\includegraphics[width=1\columnwidth]{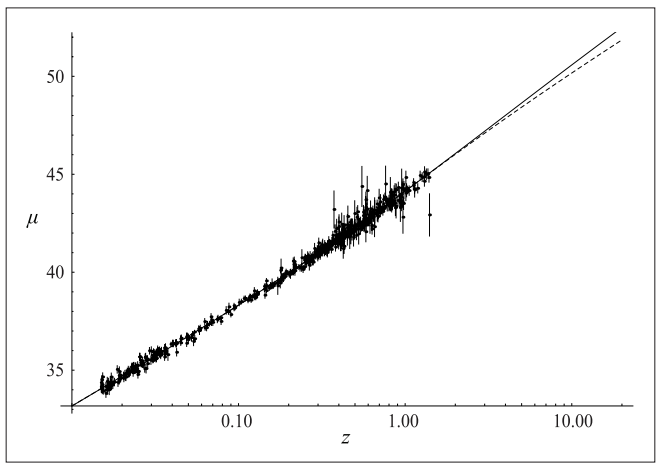}
\caption{Best fits of the type Ia supernova distance moduli $\mu$ in the Union2 compilation \cite{Amanu10}. The solid line corresponds to the best fit upper branch, the dashed line corresponds to the best fit $\Lambda$CDM model or the lower branch.}\label{SN}
\end{figure}

\begin{figure}
\includegraphics[width=1\columnwidth]{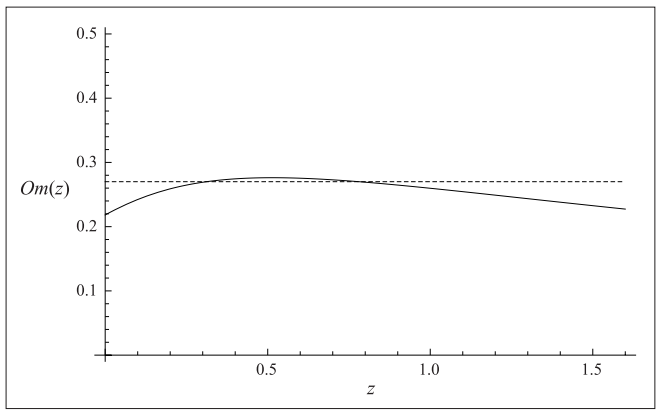}
\caption{The $Om(z)$ diagnostic of the upper branch (solid line) in comparison with the one of the $\Lambda$CDM model or the lower branch (dashed line).}\label{Om}
\end{figure}

\vskip .2cm

\noindent{\em Upper branch.} Considering $H_+(a)$, the effective cosmological constant at the fixed point turns out to be determined by the free parameters $H_*$ and $C_3$, whereas the physical interpretation of the remaining terms is far from obvious. In particular, there is no contribution from quantum fields behaving like (dark) matter or radiation. However, the upper branch solution describes the late-time evolution of the Universe at least as well as the concordance model does. In order to demonstrate this, we consider $H_+$ in terms of the redshift $z=a^{-1}-1$ and perform a $\chi^2$-fit of the latest  type Ia supernova data, the Union2 compilation \cite{Amanu10}, {\it cf.} figure \ref{SN}. To emphasise the fundamental deviation of the upper branch from the $\Lambda$CDM model, we set $C_3=0$ by hand and find the best-fit $\chi^2$ to be as low as the one of the $\Lambda$CDM model, namely, $\chi_\text{min}^2/dof\sim 0.98$. The best fit parameters are $H_*=0.570 H_0\;,\quad C_1=-0.273 H_0^4\;,\quad C_2 = 0.270 H_0^4$, where $H_0\doteq H_+(z=0)$.
The discrepancy from the predictions of $\Lambda$CDM can be made evident if we compare the two different models by means of the $\text{\it Om}(z)\doteq [(H(z)/H_0)^2-1]/[(1+z)^3-1]$ diagnostic recently introduced in \cite{Sahni}, see figure \ref{Om}. This observable is particularly well-suited to determine whether or not dark energy is a cosmological constant at low redshifts.

To summarise, the cosmological solutions associated with the upper branch provide a genuinely quantum theoretical realisation of dynamical dark energy which can easily be distinguished from a cosmological constant by upcoming cosmological observations. However, it remains unclear how to reconcile the cosmological evolution with the standard picture of matter and radiation domination in earlier stages of cosmic history.



\vskip .2cm

\noindent{\em Lower branch.} Let us now consider the class of solutions dubbed $H_-(a)$ in \eqref{effective}. For sufficiently large values of $H_*$, we can expand the square root to approximate $H^2_-(a)$ as $K_1 + K_2 a^{-3}+K_3 a^{-6}$, where the constants $K_i$ are specified in terms of $C_2$, $C_3$ and $H_*$ and where we have set $C_1=0$. Since the value of $K_3$ turns out to be of order $10^{-3}H_0^2$ in the best fit result, the model provides an effective description of the concordance model, reproducing all the observational features of $\Lambda$CDM in the recent past. However, deviations emerging from the higher order corrections are to be expected in the $z\gg1$ regime, which is not yet observationally accessible. Moreover, the fit implies that $C_2$, which equals $(\sum^{N_0}_{i=1}m_i+3\sum^{N_{1/2}}_{j=1}m_j)T^3$ up to ${\mathcal O}(1)$ numerical factors, must be of the same order as the critical density $3H^2_0(8\pi G)^{-1}$. The fit is not sufficient to fix the masses and $T$ at the same time. However, if we assume a single mass parameter $m$ for simplicity, we get $m=10\,\mathrm{eV}\left(\frac{1\textrm{K}}{T}\right)^3$.

Intriguingly, the dark energy component slightly deviates from being a pure renormalization constant. Hence, as on the upper branch, it has a dynamical character. Even more importantly, we find that a component of the underlying quantum system, namely, the $K_2 a^{-3}$ contribution to the zero-point energy of a state which fulfils an approximate KMS condition, shares the redshift behaviour of dark matter. Since this component scales with $T^3$ instead of an inverse volume factor, it does not allow for an interpretation in terms of a mass density of non-relativistic weakly interacting particles. So far, the thermal   nature of the quantum state seems to affect only the background evolution. It remains to be checked, whether its contribution to the energy density is able to reproduce also the clustering properties of dark matter inferred from astrophysical observations. As a first check, we  computed  $\langle\wick{\phi^2}\rangle_\Omega$ on a spherically symmetric and static spacetime. We estimated the corresponding lowest order thermal contribution to the vacuum energy density to be proportional to $|g_{00}|^{-3/2}$, where $g_{00}$ is the coefficient of the metric in the static time direction. Since $g_{00}$ can be approximated as $-(1+c r^2)$  inside of a spherical body \cite{Wald84}, $r$ being the radius, we infer a density profile which interpolates between a constant behaviour in the innermost region and a decay with $r^{-3}$ for large distances from the centre. This result stunningly agrees with astrophysical estimates of dark matter haloes of dwarf spheroidal galaxies \cite{Burk95, Gil07}.

\vskip .2cm

\noindent{\em Outlook.} We have applied a quantization scheme which is intrinsically suited for curved backgrounds to cosmology. Under the assumption that the underlying quantum state fulfils an approximate KMS condition at some point in the past, we have shown that there exist homogeneous and isotropic solutions of the semiclassical Einstein equations which are asymptotically stable. The exact behaviour of the Hubble function depends on suitable renormalization parameters, intrinsic to the quantization procedure, which we determined by fitting type Ia supernova data. Interestingly, there are two classes of solutions, which are both physically acceptable {\it a priori}, and which both provide a dynamical interpretation of dark energy that reproduces the observed recent expansion history of the Universe (at least) as good as the $\Lambda$CDM model.

There are still several open questions to address in future research and here we shall mention only the most relevant ones. The first, as already briefly discussed, calls us to clarify to what extent the thermal nature of quantum states entails an interpretation in terms of a new, additional contribution to the dark matter component in the Universe. In order to check whether a full quantum description of the complete dark matter sector is conceivable, it will be mandatory to study the inhomogeneous fluctuations of the thermal quantum energy density and their impact on the formation and growth of large scale structures. We also mention the necessity first to include interactions and, then, to specify a concrete model. These subsequent steps will bring us closer to a more fundamental understanding of the dynamics of the Universe.

\vskip .2cm

\begin{acknowledgments}
C.D. and T.P.H. gratefully acknowledge financial support from the DFG through the Emmy Noether Grant WO 1447/1-1, and the research clusters SFB676 and LEXI ``Connecting Particles with the Cosmos''. The work of N.P. is supported in part by the ERC Advanced Grant 227458 OACFT.  It is a pleasure to thank W. Buchm\"uller, K. Fredenhagen, and M. Wohlfarth for illuminating discussions. We are grateful to C. Hambrock for providing us with a pr\^et-\`a-porter fitting routine.
\end{acknowledgments}


\end{document}